# Visualizing Pair-breaking Scattering Interference in Bulk FeSe

Matthew Toole[1,2][†], Nileema Sharma[1,2][†], James McKenzie[1,2], Fangjun Cheng[1,2], Sheng Ran[3], and Xiaolong Liu[1,2]*

[1]Department of Physics and Astronomy, University of Notre Dame, Notre Dame, IN 46556, USA
[2]Stavropoulos Center for Complex Quantum Matter, University of Notre Dame, Notre Dame, IN 46556, USA
[3]Department of Physics, Washington University in St. Louis, St. Louis, MO 63130, USA

**Spatially periodic modulations of the superconducting gap have been recently reported in diverse materials and are often attributed to pair density wave order. An alternative mechanism, termed pair-breaking scattering interference (PBSI), was proposed to produce gap modulations without finite-momentum pairing. Here we investigate signatures of PBSI in bulk FeSe using scanning tunneling microscopy with superconductive tips, enabling enhanced energy resolution and Josephson tunneling. Subsurface magnetic scatterers with Yu–Shiba–Rusinov states are identified in FeSe, around which we observe particle–hole symmetric gap modulations accompanied by spatial modulation of the Josephson current. Those modulations have wavevectors consistent with intra-pocket PBSI. We further demonstrate that phase-referenced quasiparticle interference imaging offers an independent and direct probe of PBSI beyond gap mapping. These results establish PBSI as a viable origin of gap modulations in superconductors lacking preexisting charge/spin density wave orders, and motivate further investigation of the intriguing gap modulation phenomenology.**

Superconductivity is characterized by a complex order parameter whose magnitude defines the superconducting gap $\Delta$. Such a gap can be determined from two coherence peaks in the quasiparticle excitation spectra that are located at energies of $E^\pm = \pm\Delta^\pm$ with particle-hole symmetry ($\Delta^- = \Delta^+ = \Delta_0$). Although conventional superconductivity exhibits translational invariant $\Delta_0$, recent research has been tightly focused on scenarios where $\Delta^\pm$ are spatially modulating in energy in a periodic fashion. The most prominent state of matter that can lead to such modulated gaps is the pair density wave (PDW) state where electrons pair with a finite center-of-mass momentum $\boldsymbol{Q}$ [1]. For a PDW, the modulation of $\Delta^\pm$ is particle-hole symmetric such that $\Delta^-(\boldsymbol{r}) = \Delta^+(\boldsymbol{r}) \propto \cos(\boldsymbol{Q} \cdot \boldsymbol{r})$. While a fundamental PDW can exist in the absence of uniform superconductivity, for example, via mechanisms favoring intra-pocket pairing away from the $\Gamma$ point [2,3] or in a strong magnetic field [4,5], a composite PDW can also arise when uniform superconductivity couples to a preexisting spin or charge density wave (S/CDW), producing secondary gap and superfluid density modulations [6].

Phenomenologically, regardless of the mechanism and modulation length scale, a suite of materials is recently demonstrated to host particle-hole symmetric modulations of $\Delta^\pm$ [7-27] in scanning tunneling microscopy (STM) measurements, consistent with a PDW or pair density modulation [23-25] state. Furthermore, the associated modulation in superfluid density has been directly visualized using scanned Josephson tunneling microscopy (SJTM) [7,10,11,18,19,25]. In superconductors with a dominant S/CDW, particle-hole symmetric gap modulations (up to a few tens of μV) can be reasonably attributed to the formation of composite PDWs [11,12,13,18,19,20,23]. However, for superconductors without any preexisting density wave orders, a recent theoretical proposal [28] (initially introduced in the context of cuprates [29]) introduces an alternative mechanism other than PDW that can lead to periodic gap modulations, both in particle-hole symmetric (like PDW) and asymmetric fashions. The mechanism, coined pair-breaking scattering interference (PBSI), cautions against the assertion of fundamental PDWs when no S/CDW exist in the normal state.

Motivated by the need to distinguish genuine PDWs from alternative mechanisms, we report experimental signatures of PBSI in bulk FeSe, which is chosen for several reasons. Firstly, many of the modulated superconducting states [15,21,22,23,24,25] have been observed in the family of Fe-based superconductors, among which FeSe has the simplest structure with well-established Fermi surface and gap structures. Secondly, no spin or charge order exists in bulk FeSe such that no composite PDWs are expected. No fundamental PDWs have been observed in FeSe either. Thirdly, the absence of chemical, mechanical, and interfacial heterogeneity in bulk FeSe allows the signatures of PBSI to be isolated and examined with the highest level of clarity. We use SJTM at 300 mK with superconductive Nb tips, ensuring maximal energy resolution beyond the Fermi Dirac limit [8,11,13,16,18] for resolving subtle gap modulations. Two sets of coherence peaks associated with the anisotropic gaps of the electron and hole pockets of FeSe are clearly resolved, which individually exhibit particle-hole symmetric spatial modulations in orthogonal directions with wavevectors consistent with the intra-pocket PBSI mechanism. As a consequence of the gap modulations, we further propose and detect signatures of PBSI based on phase-referenced differential conductance $[g(\mathbf{r}, E = eV) \equiv dI/dV]$ imaging, which offers a more convenient experimental probe than direct gap mapping. Our results demonstrate that gap modulations can indeed arise purely from PBSI in a density-wave-free superconductor, establishing a benchmark for distinguishing PBSI from true finite-momentum pairing.

*Corresponding author: xliu33@nd.edu

[†]These authors contributed equally to this work

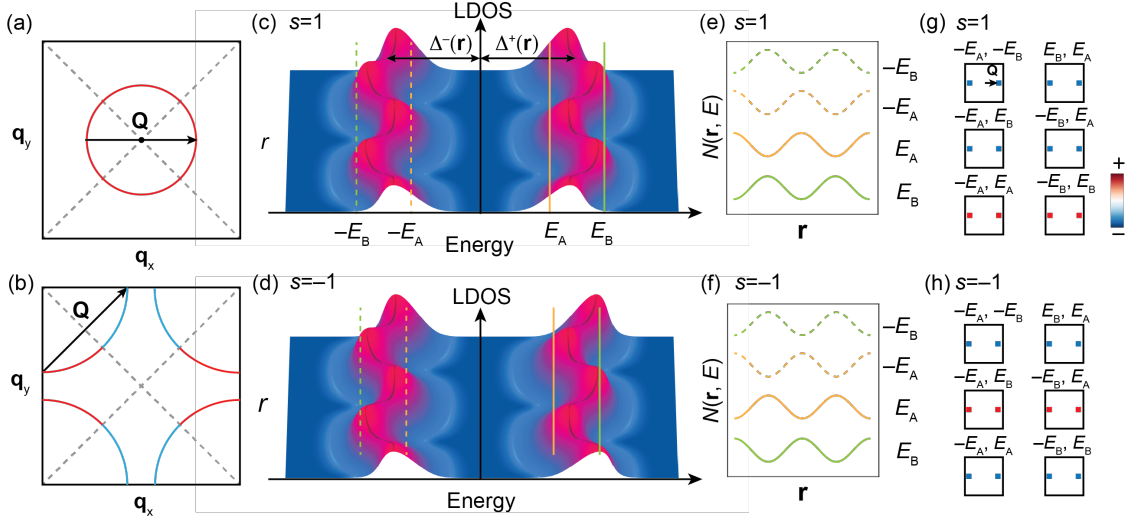

FIG 1. (a) Schematic of a PBSI vector $\mathbf{Q}$ connecting portions of the normal state Fermi surface with the same and (b) opposite signs of the superconducting order parameter. They correspond to $s = \pm 1$, respectively. (c) LDOS as a function of position showing particle-hole symmetric and (d) asymmetric modulations for $s = \pm 1$, respectively. (e,f) LDOS as a function of position taken at four different energies as indicated by the dashed and solid lines in panels (c,d) for $s = \pm 1$, respectively. (g,h) Anticipated $g_{PR}(\mathbf{q}, E_1, E_2)$ signals for $s = \pm 1$, respectively, for $E_1, E_2 \in \{\pm E_A, \pm E_B\}$. The sign of the hot spot at $\mathbf{Q}$ depends on $s$ and is color coded.

In a superconductor with an anisotropic gap $\Delta(\mathbf{k})$, Bogoliubov quasiparticles inherit the phase of the order parameter during scattering processes. Such scattering predominantly takes place between the tips of banana-shaped constant energy contours (CECs) in the momentum space, which have the highest density of states [30]. If the portions of $\Delta(\mathbf{k})$ connected by the scattering vector $\mathbf{Q}$ have the same sign ($s = 1$) as shown in Fig. 1(a), the scattering is pair breaking only in the presence of magnetic scatterers. On the other hand, if the scattering connects sign-changing ($s = -1$) portions of $\Delta(\mathbf{k})$, pair breaking takes place even with non-magnetic scatterers as shown in Fig. 1(b), where blue and red colors denote opposite signs of $\Delta$.

According to the PBSI formulism [28], the local density of states (LDOS) $N(\mathbf{r}, E)$ for $s = \pm 1$ are periodically modulated as

$N^{s=1}(\mathbf{r}, E) = [1 - W(h) \cos(\mathbf{Q} \cdot \mathbf{r})][\delta(E + \Delta_S) + \delta(E - \Delta_S)] + [1 + W(h) \cos(\mathbf{Q} \cdot \mathbf{r})][\delta(E + \Delta_L) + \delta(E - \Delta_L)]$ (1)

$N^{s=-1}(\mathbf{r}, E) = [1 - \cos(\mathbf{Q} \cdot \mathbf{r})][\delta(E + \Delta_L) + \delta(E - \Delta_S)] + [1 + \cos(\mathbf{Q} \cdot \mathbf{r})][\delta(E + \Delta_S) + \delta(E - \Delta_L)]$ (2)

Here, $\Delta_{L,S} = \Delta_0 \pm V$, $V$ is the scattering potential, $\Delta_0$ is the average gap, and $W(h)$ is a monotonic function of Zeeman interactions $h$ with $W(0) = 0$. Consequently, the coherence peaks at $E^\pm(\mathbf{r}) = \pm \Delta^\pm(\mathbf{r})$ modulate between $\pm \Delta_S$ and $\pm \Delta_L$ in particle-hole symmetric (for $s = 1$) and asymmetric (for $s = -1$) fashions as shown in Figs. 1(c,d). For $s = 1$, such gap modulations will exist as long as $W(h) \neq 0$, e.g., in the presence of magnetic impurities. However, the amplitude of the gap modulation ($V$) is independent of the Zeeman interaction strength according to eqn. (1).

The energy- and $s$-dependent modulations of $N^{s=\pm 1}(\mathbf{r}, E)$ are schematically shown in Figs. 1(e,f) at four energies ($\pm E_A, \pm E_B$) that are inside and outside the two coherence peaks. While modulations between $\pm E_A$ and $\pm E_B$ are always out of phase, modulations between $E_{A,B}$ and $-E_{A,B}$ are in and out of phase for $s = \pm 1$, respectively. More quantitatively, we can define a generalized phase-referenced quasiparticle interference (QPI) signal [31,32] in the Fourier space as $g_{PR}(\mathbf{q}, E_1, E_2) = |g(\mathbf{q}, E_1)| \cos[\theta(\mathbf{q}, E_1) - \theta(\mathbf{q}, E_2)]$, where $\theta(\mathbf{q}, E)$ is the phase of the Fourier transform $g(\mathbf{q}, E)$ at $\mathbf{q}$. Then, an in- and out-of-phase $N(\mathbf{r}, E) \propto g(\mathbf{r}, E)$ modulations at $E_1$ and $E_2$ will be manifested as positive and negative peaks at the modulation wavevector $\mathbf{Q}$ in $g_{PR}(\mathbf{q}, E_1, E_2)$, respectively. For $E_1, E_2 \in \{\pm E_A, \pm E_B\}$, the expected $g_{PR}(\mathbf{q}, E_1, E_2)$ is schematically given in Figs. 1(g,h) for $s = \pm 1$ with the sign denoted by color, respectively. Because it is generally more straightforward to acquire energy-dependent $g(\mathbf{r}, E)$ images than gap maps, $g_{PR}(\mathbf{q}, E_1, E_2)$ provides another independent method to detect and diagnose PBSI-induced gap modulations.

*Corresponding author: xliu33@nd.edu
†These authors contributed equally to this work

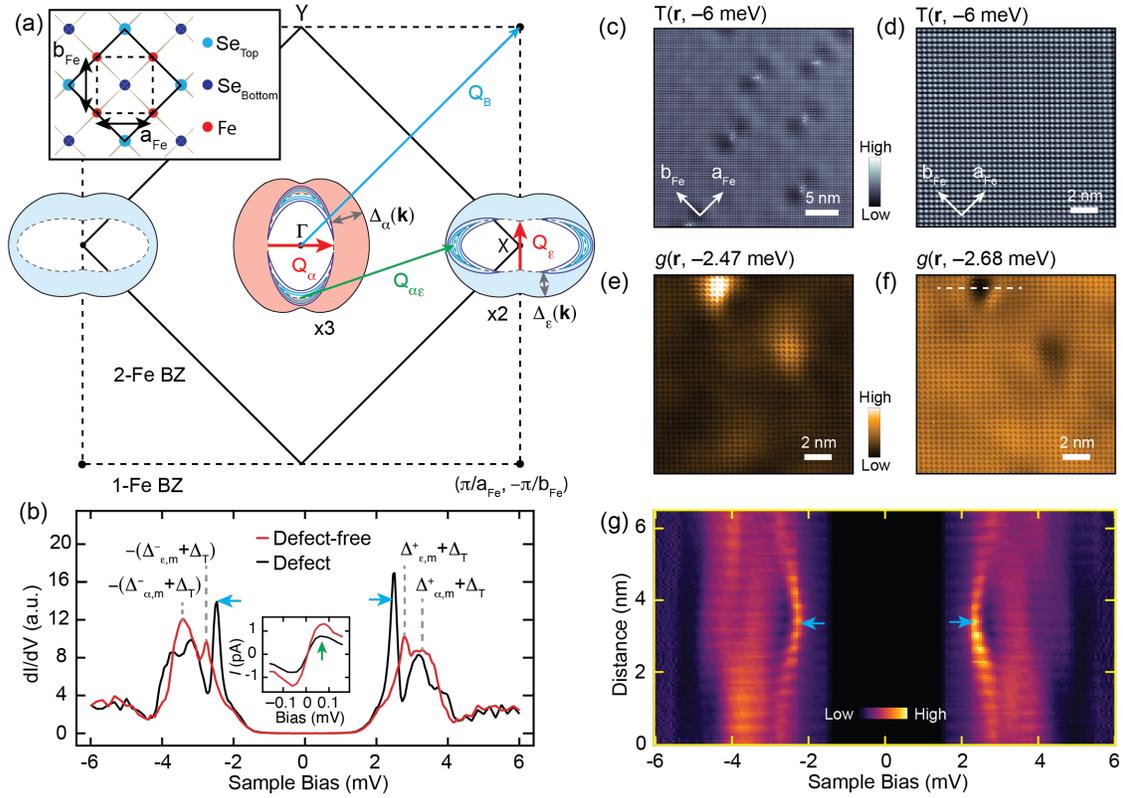

FIG 2. (a) Schematic of the superconducting gap structures $\Delta_{\alpha,\varepsilon}(\mathbf{k})$ and signs of the order parameters on the anisotropic $\alpha$ and $\varepsilon$ pockets, which are enlarged by a factor of three and two, respectively, for clarity. The banana-shaped sub-gap CECs are shown with colors encoding their energy (darker blue indicates higher energy). $\mathbf{Q}_{\alpha,\varepsilon}$, $\mathbf{Q}_{\alpha\varepsilon}$ denote the intra-pocket and inter-pocket scattering vectors, respectively. $\mathbf{Q}_B$ is a Bragg vector. The grey dashed lines indicate the normal state Fermi surfaces of the $\alpha$ and $\varepsilon$ pockets. Inset: FeSe crystal structure viewed from the out-of-plane direction. (b) $dI/dV$ spectra taken away from and at a subsurface impurity using a superconductive Nb tip at 0.3 K. The two sets of coherence peaks are located at $\pm(\Delta^{\pm}_{\alpha,m} + \Delta_T)$ and $\pm(\Delta^{\pm}_{\varepsilon,m} + \Delta_T)$. The in-gap states from the impurity are indicated by the blue arrows. Inset: Josephson tunneling current spectra near zero bias voltage taken away from and at a subsurface impurity. (c) Topographic image of a single nematic domain of FeSe, exhibiting dumbbell-shaped defects that are elongated along the $a_{Fe}$ direction. (d) A topographic image of a surface-defect-free region and the associated (e) $g(\mathbf{r}, -2.47 \text{ meV})$ and (f) $g(\mathbf{r}, -2.68 \text{ meV})$ images showing the existence of two subsurface defects. (g) A series of $dI/dV$ spectra taken along the dashed line in panel (f) across one of the defects, revealing spatially dispersing in-gap bound states as indicated by the blue arrows.

FeSe is a nematic $s^{\pm}$-wave superconductor with orbital-selective electron pairing from the $d_{yz}$ orbitals of Fe. We denote the superconducting gaps on the $\alpha$ (at $\Gamma$) and $\varepsilon$ (at X) pockets as $\Delta_{\alpha,\varepsilon}(\mathbf{k})$, respectively (Fig. 2(a)), which are manifested as two sets of coherence peaks in the d$I$/d$V$ spectra at $\pm(\Delta^{\pm}_{\alpha,m} + \Delta_T) \approx \pm 3.5$ meV and $\pm(\Delta^{\pm}_{\varepsilon,m} + \Delta_T) \approx \pm 2.7$ meV as shown by the spectrum in red in Fig. 2(b). Here, $\Delta^{\pm}_{\alpha,m} \approx 2.3$ meV and $\Delta^{\pm}_{\varepsilon,m} \approx 1.5$ meV are the gap maxima on the α and ε pockets of FeSe [30], respectively, and $\Delta_T \approx 1.2$ meV is the gap of Nb tip. As a result of **k**-dependent orbital characters of the Fermi surfaces, both gaps are highly anisotropic, leading to banana-shaped sub-gap CECs shown in Fig. 2(a) (darker blue indicates higher energy). The dominant sign-preserving intra-pocket PBSI vectors are therefore $\mathbf{Q}_{\alpha,\varepsilon}$ in the α and ε pocket, respectively. The Se lattice Bragg vector ($\mathbf{Q}_B$, Fig. 2(a)) can also be considered as a sign-preserving inter-pocket scattering vector, although less significant because gap modulations with the atomic lattice is always expected given the periodic Bloch wavefunctions of any crystal. The inter-pocket scattering vector connecting portions of the α and ε pockets with identical orbital characteristics is experimentally detectable in phase-referenced QPI imaging and show unique signatures of the sign-changing gaps (Fig. S1 [33]). However, because the two pockets possess vastly different gap functions resulting in two sets of coherence peaks, the PBSI formulism [28] involving a single gap (i.e., intra-pocket) cannot be directly applied. Therefore, in this study, we focus on detecting evidence of intra-pocket PBSI at $\mathbf{Q}_{\alpha,\varepsilon}$ near magnetic impurities. A typical topographic image of FeSe in a single nematic domain is shown in Fig. 2(c), where the $a_{Fe}$-axis can be determined by the directional alignment of the dumbbell-shaped defects. In a surface-defect-free region (Fig. 2(d)), two subsurface point defects are revealed in $g(\mathbf{r}, E)$ imaging in Figs. 2(e,f), where they appear bright and dark, respectively (more energy-dependent images are given in Fig. S2 [33]). As shown in Fig. 2(b), the d$I$/d$V$ spectrum taken on the defect exhibits two sharp in-gap states at $E^{\pm}_D \approx \pm 2.47$ meV but with asymmetric peak height (blue arrows). This is reminiscent of the Yu-Shiba-Rusinov (YSR) states caused by magnetic impurities [34]. Moreover, by taking a series of d$I$/d$V$ spectra across one of the defects as indicated by the dashed line in Fig. 2(f), the defect-bound-states are found to be spatially dispersing in a particle-hole symmetric fashion before merging with $\pm(\Delta^{\pm}_{\varepsilon,m} + \Delta_T)$. This is also consistent with the recently reported spatially dispersing YSR states in Fe(Se,Te) [35]. Using SJTM, we further reveal a lower Josephson current at the defect (inset of Fig. 2(b)), again consistent with it being a magnetic impurity.

*Corresponding author: xliu33@nd.edu

†These authors contributed equally to this work

Next, we perform gap mapping in the same field of view (FOV) by acquiring d$I$/d$V$ spectrum at each imaging pixel to directly search for the gap modulation signature of PBSI. Compared to metallic tips, the superconductive Nb tip gaps out the thermal smearing at the Fermi level, drastically improving the energy resolution [8,11,13,16,18,36] (Fig. S3 [33]). To extract $\Delta^{\pm}_{\alpha,m}(\mathbf{r})$ and $\Delta^{\pm}_{\varepsilon,m}(\mathbf{r})$, we deconvolute d$I$/d$V$ spectra measured in the FeSe-Nb junction to obtain the LDOS of FeSe, which is then fitted with realistic twofold symmetric gaps for both the α and ε pockets [36] (Supplemental Material Sections II, III, Fig. S4 [33]). As shown in Figures 3(a-d), periodic modulations of $\Delta^{\pm}_{\alpha,m}(\mathbf{r})$ are present in real space, which are better visualized in the magnitudes of their Fourier transforms (Figs. 3(b,d)) as indicated by the red arrows. Both gap modulations correspond to a wavevector pointing along the $\mathbf{q}_{a_{Fe}}$ direction, consistent with the intra-pocket scattering vector $\mathbf{Q}_{\alpha}$ (Fig. 2(a)). To explore the spatial phase correlations of $\Delta^{\pm}_{\alpha,m}(\mathbf{r})$, we examine the symmetrized ($\Delta^S_{\alpha} \equiv \Delta^-_{\alpha,m} + \Delta^+_{\alpha,m}$) and anti-symmetrized ($\Delta^{AS}_{\alpha} \equiv \Delta^-_{\alpha,m} - \Delta^+_{\alpha,m}$) gaps in Figs. 3(e-h). Clearly, while the intensity of the gap modulation wavevector is enhanced in $\Delta^S_{\alpha}(\mathbf{q})$, it is suppressed in $\Delta^{AS}_{\alpha}(\mathbf{q})$, indicating in-phase spatial modulations of $\Delta^{\pm}_{\alpha,m}(\mathbf{r})$. Quantitatively, as shown in Fig. 3(i) by taking line profiles along the $\mathbf{q}_{a_{Fe}}$ direction in Figs. 3(b,d,f,h), the gap modulation wavevector ($1.3$ nm$^{-1}$) closely matches that expected for the intra-α-pocket scattering vector $\mathbf{Q}_{\alpha}$ determined from QPI imaging [30]. To directly visualize the gap modulations, we show in Fig. 3(j) a series of deconvoluted FeSe's LDOS along the yellow dashed line in Fig. 3(a). The extracted coherence peak positions are indicated by blue circles (Supplemental Material Sections III [33]), which clearly show particle-hole symmetric modulations with a wavelength of $2\pi/Q_{\alpha}$ and amplitude of 0.12 meV. Because of the finite spatial averaging effects, this gives a measure of the lower bound of the pair-breaking impurity scattering potential $V \geq 0.12$ meV, suggesting very weak scattering. Similar results are obtained for $\Delta^{\pm}_{\varepsilon,m}$ (Fig. S5 [33]), which modulate in a particle-hole symmetric fashion with a wavevector matching the intra-ε-band $\mathbf{Q}_{\varepsilon}$ along the $\mathbf{q}_{b_{Fe}}$ direction. Although present, the signal is less clear compared to $\mathbf{Q}_{\alpha}$ possibly because of the smaller wavevector and it is more favorable for electrons to tunnel into the α pocket than the ε pocket due to the smaller in-plane momentum associated with the α pocket, leading to a larger tunneling decay length [37]. Therefore, we have demonstrated the existence of intra-pocket particle-hole symmetric modulations of the superconductivity gaps in both the α and ε pockets with the anticipated wavevectors, consistent with predictions of the PBSI formulism for $s = 1$. Using SJTM, we further imaged the maximum Josephson tunneling current distribution ($I_m \propto I_J^2$, where $I_J$ is the Josephson critical current [7]) as shown in Fig. 3(l) around one of the magnetic impurities (Fig. 3(k)). In addition to a lower $I_m(\mathbf{r})$ around the impurity site (red dot), evidence of $I_m(\mathbf{r})$ modulations along the $\mathbf{q}_{a_{Fe}}$ direction is faintly observed. This is indicated by the white dashed lines in Fig. 3(l) placed on top of the crests of such modulations, consistent with theoretical predictions of simultaneous $I_m(\mathbf{r})$ and $\Delta(\mathbf{r})$ modulations around magnetic impurities [38].

*Corresponding author: xliu33@nd.edu

†These authors contributed equally to this work

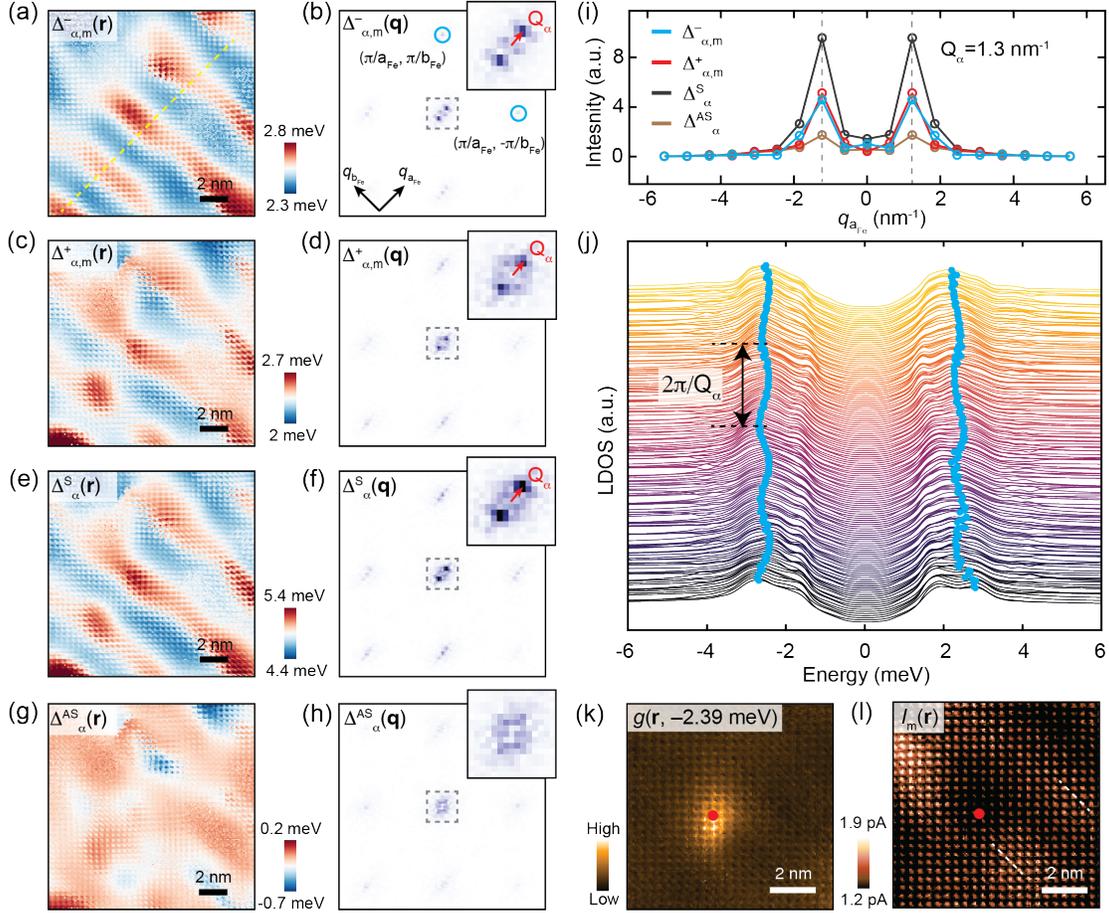

FIG 3. (a) Atomically resolved $\Delta^-_{\alpha,m}(\mathbf{r})$ and (b) the magnitude of its Fourier transform. The Se lattice Bragg peaks are indicated by the blue circles. The inset shows a zoomed-in view of the low-$\mathbf{Q}$ region indicated by the dashed square, which shows clearly the gap modulation wavevector corresponding to $\mathbf{Q}_\alpha$. (c) Atomically resolved $\Delta^+_{\alpha,m}(\mathbf{r})$ and (d) the magnitude of its Fourier transform. (e) Atomically resolved $\Delta^S_\alpha(\mathbf{r})$ and (f) the magnitude of its Fourier transform. (g) Atomically resolved $\Delta^{AS}_\alpha(\mathbf{r})$ and (h) the magnitude of its Fourier transform. While $\mathbf{Q}_\alpha$ is enhanced in $\Delta^S_\alpha(\mathbf{r})$, it is strongly suppressed in $\Delta^{AS}_\alpha(\mathbf{r})$. (i) Line profiles of the Fourier transform magnitudes in panels (b,d,f,h) along the $\mathbf{q}_{a_{Fe}}$ direction. (j) A series of LDOS spectra of FeSe taken along the dashed line in panel (a), showing the coherence peak (blue circles) modulations of $\Delta^\pm_{\alpha,m}(\mathbf{r})$. (k) $g(\mathbf{r}, -2.39\ \text{meV})$ image and (l) maximum Josephson current $I_m(\mathbf{r})$ image around a subsurface magnetic impurity indicated by the red dot. The dashed lines in panel (l) are located at the crests of $I_m(\mathbf{r})$ modulations.

Finally, we examine the signatures of PBSI independently from $g_{PR}(\mathbf{q}, E_1, E_2)$ as we proposed in Figs. 1(g,h). Figures 4(a-d) show simultaneously acquired $g(\mathbf{r}, E)$ at two pairs of energies, $\pm 3.87$ meV and $\pm 3.23$ meV, which are outside and inside the pair of coherence peaks at $\pm(\Delta^\pm_{\alpha,m} + \Delta_T) \approx \pm 3.5$ meV. In-phase periodic modulations of $g(\mathbf{r}, E)$ are seen within each pair (Figs. 4(a,b), Figs. 4(c,d)); however, the modulations between $g(\mathbf{r}, \pm 3.87\ \text{meV})$ and

*Corresponding author: xliu33@nd.edu

†These authors contributed equally to this work

$g(\mathbf{r}, \pm 3.23 \text{ meV})$ are out of phase. These can be seen from the blue lines, which are positioned at the crests of $g(\mathbf{r}, \pm 3.87 \text{ meV})$, but at the valleys of $g(\mathbf{r}, \pm 3.23 \text{ meV})$. In Fig. 4(e), comparing the line profiles along the dashed line in Fig. 4(a) further confirms this observation. Quantitatively and by taking into account the entire FOV, we examine the signs of $g_{PR}(\mathbf{q}, E_1, E_2)$ in Fig. 4(f) for $E_1, E_2 \in \{\pm 3.23 \text{ meV}, \pm 3.87 \text{ meV}\}$. All $g_{PR}(\mathbf{q}, E_1, E_2)$ images exhibit a hot spot that is consistent with $\mathbf{Q}_\alpha$, but with varying signs that are consistent with predictions in Fig. 1(g) for $s=1$. Furthermore, as shown in Fig. S6 [33], the measured relative phases of modulation between $g(\mathbf{r}, E)$ and $\Delta^\pm_{\alpha/\varepsilon,m}(\mathbf{r})$ at $\mathbf{Q}_{\alpha/\varepsilon}$ agree with PBSI-induced co-modulations of LDOS and gap (Fig. 1(c)), demonstrating the internal consistency between $\Delta^\pm_{\alpha/\varepsilon,m}(\mathbf{r})$ and $g(\mathbf{r}, E)$ measurements.

In summary, we have identified subsurface magnetic impurities that give rise to spatially dispersing in-gap YSR states in bulk FeSe. Through gap mapping with drastically improved energy resolution using superconductive tips, Josephson current imaging, and phase-referenced QPI visualization, we have observed experimental evidence consistent with the recently proposed PBSI formalism for intra-pocket scattering in bulk FeSe. Importantly, the observed gap modulation amplitude of 0.12 meV is substantially larger than those in materials with composite PDWs [11,12,13,18,19,20], underscoring the need for caution when attributing periodic and sizable gap modulations to PDWs when there are no other dominant density wave orders. Our study motivates further theoretical work to elucidate PBSI signatures arising from inter-pocket scattering processes involving multiple anisotropic superconducting gaps, as well as the understanding of superconducting gap modulations at various length scales.

*Note added.* Recently, we became aware of another work by Hou et al. [39], reporting atomic-scale gap modulations in FeTe$_{0.55}$Se$_{0.45}$ attributed to PBSI.

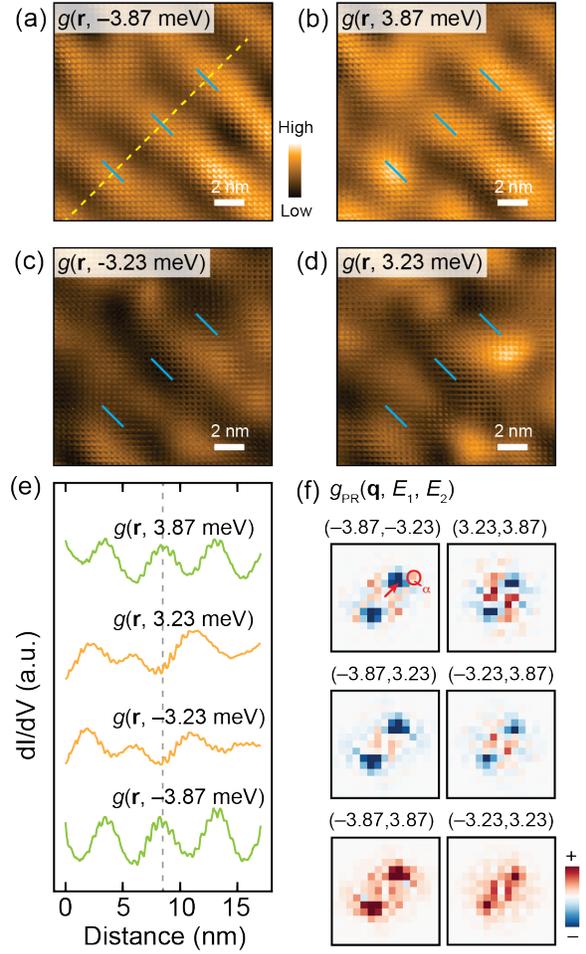

FIG 4. (a) $g(\mathbf{r}, -3.87 \text{ meV})$, (b) $g(\mathbf{r}, 3.87 \text{ meV})$, (c) $g(\mathbf{r}, -3.23 \text{ meV})$, and (d) $g(\mathbf{r}, 3.23 \text{ meV})$ images of the same FOV as in Fig. 3. The blue lines are at the same locations in all cases, showing an out-of-phase modulation between $g(\mathbf{r}, \pm 3.87 \text{ meV})$ and $g(\mathbf{r}, \pm 3.23 \text{ meV})$. (e) Line profiles of $g(\mathbf{r}, \pm 3.87 \text{ meV})$ and $g(\mathbf{r}, \pm 3.23 \text{ meV})$ taken along the dashed line in panel (a). (f) Phase-referenced QPI signal $g(\mathbf{q}, E_1, E_2)$ for $E_1, E_2 \in \{\pm 3.23 \text{ meV}, \pm 3.87 \text{ meV}\}$. We only need to examine these six combinations of $E_1$ and $E_2$ because $sign[g(\mathbf{q}, E_1, E_2)] = sign[g(\mathbf{q}, E_2, E_1)]$.

### ACKNOWLEDGMENTS

The authors thank A. Kreisel and A. Kostin for helpful communications. X.L. acknowledges support from the National Science Foundation Division of Materials Research Award DMR-2443396. N.S. and M.T. acknowledge support from the Notre Dame Materials Science and Engineering Fellowship. S.R. acknowledges support from the National Science Foundation Division of Materials Research Award DMR-2512965.

*Corresponding author: xliu33@nd.edu

†These authors contributed equally to this work

[1] D. F. Agterberg, J. C. Séamus Davis, S. D. Edkins, E. Fradkin, D. J. Van Harlingen, S. A. Kivelson, P. A. Lee, L. Radzihovsky, J. M. Tranquada, and Y. Wang, The physics of pair-density waves: cuprate superconductors and beyond, Annu. Rev. Condens. Matter Phys. **11**, 231 (2020).
[2] Y.-T. Hsu, A. Vaezi, M. H. Fischer, and E.-A. Kim, Topological superconductivity in monolayer transition metal dichalcogenides, Nat. Commun. **8**, 14985 (2017).
[3] C. Yoon, T. Xu, Y. Barlas, and F. Zhang, Quarter metal superconductivity, arXiv:2502.17555.
[4] J. J. Kinnunen, J. E. Baarsma, J.-P. Martikainen, and P. Törmä, The Fulde–Ferrell–Larkin–Ovchinnikov state for ultracold fermions in lattice and harmonic potentials: a review, Rep. Prog. Phys. **81**, 046401 (2018).
[5] S. D. Edkins, A. Kostin, K. Fujita, A. P. Mackenzie, H. Eisaki, S. Uchida, S. Sachdev, M. J. Lawler, E.-A. Kim, J. C. Séamus Davis, and M. H. Hamidian, Magnetic field–induced pair density wave state in the cuprate vortex halo, Science **364**, 976 (2019).
[6] A. Laskowski, J. Bedow, and D. K. Morr, Josephson scanning tunneling spectroscopy in superconducting phases coexisting with pair-, charge- and spin-density-waves, arXiv:2512.02191.
[7] M. H. Hamidian, S. D. Edkins, S. H. Joo, A. Kostin, H. Eisaki, S. Uchida, M. J. Lawler, E.-A. Kim, A. P. Mackenzie, K. Fujita, J. Lee, and J. C. Séamus Davis, Detection of a Cooper-pair density wave in $Bi_2Sr_2CaCu_2O_{8+x}$, Nature **532**, 343 (2016).
[8] Z. Du, H. Li, S. H. Joo, E. P. Donoway, J. Lee, J. C. Séamus Davis, G. Gu, P. D. Johnson, and K. Fujita, Imaging the energy gap modulations of the cuprate pair-density-wave state. Nature **580**, 65 (2020).
[9] W. Ruan, X. Li, C. Hu, Z. Hao, H. Li, P. Cai, X. Zhou, D.-H. Lee, and Y. Wang, Visualization of the periodic modulation of Cooper pairing in a cuprate superconductor, Nat. Phys. **14**, 1178 (2018).
[10] W. Chen, W. Ren, N. Kennedy, M. H. Hamidian, S. Uchida, H. Eisaki, P. D. Johnson, S. M. O'Mahony, and J. C. Séamus Davis, Identification of a nematic pair density wave state in $Bi_2Sr_2CaCu_2O_{8+x}$, Proc. Natl. Acad. Sci. **119**, e2206481119 (2022).
[11] X. Liu, Y. X. Chong, R. Sharma, and J. C. Séamus Davis, Discovery of a Cooper-pair density wave state in a transition-metal dichalcogenide, Science **372**, 1447 (2021).
[12] L. Cao, Y. Xue, Y. Wang, F.-C. Zhang, J. Kang, H.-J. Gao, J. Mao, and Y. Jiang, Directly visualizing nematic superconductivity driven by the pair density wave in $NbSe_2$, Nat. Commun. **15**, 7234 (2024).
[13] Q. Gu, J. P. Carroll, S. Wang, S. Ran, C. Broyles, H. Siddiquee, N. P. Butch, S. R. Saha, J. Paglione, J. C. Séamus Davis, and X. Liu, Detection of a pair density wave state in $UTe_2$, Nature **618**, 921 (2023).
[14] A. Aishwarya, J. May-Mann, A. Raghavan, L. Nie, M. Romanelli, S. Ran, S. R. Saha, J. Paglione, N. P. Butch, E. Fradkin, and V. Madhavan, Magnetic-field-sensitive charge density waves in the superconductor $UTe_2$, Nature **618**, 928 (2023).
[15] H. Zhao, R. Blackwell, M. Thinel, T. Handa, S. Ishida, X. Zhu, A. Iyo, H. Eisaki, A. N. Pasupathy, and K. Fujita, Smectic pair density-wave order in $EuRbFe_4As_4$, Nature **618**, 940 (2023).
[16] H. Chen, H. Yang, B. Hu, Z. Zhao, J. Yuan, Y. Xing, G. Qian, Z. Huang, G. Li, Y. Ye, S. Ma, S. Ni, H. Zhang, Q. Yin, C. Gong, Z. Tu, H. Lei, H. Tan, S. Zhou, C. Shen, X. Dong, B. Yan, Z. Wang, and H.-J. Gao, Roton pair density wave in a strong-coupling kagome superconductor, Nature **599**, 222 (2021).
[17] X. Han, H. Chen, H. Tan, Z. Cao, Z. Huang, Y. Ye, Z. Zhao, C. Shen, H. Yang, B. Yan, Z. Wang, and H.-J. Gao, Atomic manipulation of the emergent quasi-2D superconductivity and pair density wave in a kagome metal. Nat. Nanotechnol. **20**, 1017 (2025).
[18] H. Deng, H. Qin, G. Liu, T. Yang, R. Fu, Z. Zhang, X. Wu, Z. Wang, Y. Shi, J. Liu, H. Liu, X.-Y. Yan, W. Song, X. Xu, Y. Zhao, M. Yi, G. Xu, H. Hohmann, S. C. Holbæk, M. Dürrnagel, S. Zhou, G. Chang, Y. Yao, Q. Wang, Z. Guguchia, T. Neupert, R. Thomale, M. H. Fischer, and J.-X. Yin, Chiral kagome superconductivity modulations with residual Fermi arcs, Nature **632**, 775 (2024).
[19] X.-Y. Yan, G. Liu, H. Deng, H. Xu, H. Ma, H. Qin, J.-Y. Zhang, Y. Zhao, H. Zhao, Z. Qu, Y. Zhong, K. Okazaki, X. Zheng, Y. Peng, Z. Guguchia, X. X. Wu, Q. Wang, X.-H. Fan, W. Song, M.-W. Gao, H. Hohmann, M. Durrnagel, R. Thomale, and J.-X. Yin, Phase-sensitive evidence for 2 × 2 pair density wave in a kagome superconductor, arXiv:2510.10958.
[20] W. Song, X.-Y. Yan, X. Yu, D. Wu, D. Hu, H. Qin, G. Liu, H. Deng, C. Yan. M. Gao, Z. Wang, R. Wu, and J.-X. Yin, Switchable chiral 2 × 2 pair density wave in pure $CsV_3Sb_5$, arXiv:2510.12730.
[21] Y. Liu, T. Wei, G. He, Y. Zhang, Z. Wang, and J. Wang, Pair density wave state in a monolayer high-$T_c$ iron-based superconductor, Nature **618**, 934 (2023).
[22] T. Wei, Y. Liu, W. Ren, Z. Wang, and J. Wang, Intertwined charge and pair density orders in a monolayer high-$T_c$ iron-based superconductor, arXiv:2305.17991.
[23] L. Kong, M. Papaj, H. Kim, Y. Zhang, E. Baum, H. Li, K. Watanabe, T. Taniguchi, G. Gu, P. A. Lee, and S. Nadj-Perge, Cooper-pair density modulation state in an iron-based superconductor, Nature **640**, 55 (2025).

*Corresponding author: xliu33@nd.edu

†These authors contributed equally to this work

[24] T. Wei, Y. Liu, W. Ren, Z. Liang, Z. Wang, and J. Wang, Observation of superconducting pair density modulation within lattice unit cell, Chinese Phys. Lett. **42**, 027404 (2025).

[25] Y. Zhang, L. Yang, C. Liu, W. Zhang, and Y.-S. Fu, Visualizing uniform lattice-scale pair density wave in single-layer FeSe/SrTiO$_3$ films, arXiv:2406.05693.

[26] L.-X. Wei, P.-C. Xiao, F. Li, L. Wang, B.-Y. Deng, F.-J. Cheng, F.-W. Zheng, N. Hao, P. Zhang, X.-C. Ma, Q.-K. Xue, and C.-L. Song, Unidirectional charge and pair density waves in topological monolayer $1T'$−MoTe$_2$, Phys. Rev. B **112**, L060503 (2025).

[27] F.-J. Cheng, C.-C. Lou, A.-X. Chen, L.-X. Wei, Y. Liu, B.-Y. Deng, F. Li, Z. Wang, Q.-K. Xue, X.-C. Ma, and C.-L. Song, Imaging sublattice Cooper-pair density waves in monolayer $1T'$−MoTe$_2$, Phys. Rev. Lett. **135**, 166201 (2025).

[28] Z.-Q. Gao, Y.-P. Lin, and D.-H. Lee, Pair-breaking scattering interference as a mechanism for superconducting gap modulation, Phys. Rev. B **110**, 224509 (2024).

[29] C. Zou, Z. Hao, X. Luo, S. Ye, Q. Gao, M. Xu, X. Li, P. Cai, C. Lin, X. Zhou, D.-H. Lee, and Y. Wang, Particle–hole asymmetric superconducting coherence peaks in overdoped cuprates, Nat. Phys. **18**, 551 (2022).

[30] P. O. Sprau, A. Kostin, A. Kreisel, A. E. Böhmer, V. Taufour, P. C. Canfield, S. Mukherjee, P. J. Hirschfeld, B. M. Andersen, and J. C. Séamus Davis, Discovery of orbital-selective Cooper pairing in FeSe, Science **357**, 75 (2017).

[31] S. Chi, W. N. Hardy, R. Liang, P. Dosanjh, P. Wahl, S. A. Burke, and D. A. Bonn, Determination of the superconducting order parameter from defect bound state quasiparticle interference, arXiv:1710.09089.

[32] G. Chen, A. Aishwarya, M. R. Hirsbrunner, J. O. Rodriguez, L. Jiao, L. Dong, N. Mason, D. Van Harlingen, J. Harter, S. D. Wilson, T. L. Hughes, and V. Madhavan, Evidence for a robust sign-changing $s$-wave order parameter in monolayer films of superconducting Fe (Se,Te)/Bi$_2$Te$_3$, npj Quantum Mater. **7**, 110 (2022).

[33] See Supplemental Material at URL for details on materials, methods and additional analysis details.

[34] M. Ruby, B. W. Heinrich, Y. Peng, F. von Oppen, and K. J. Franke, Wave-function hybridization in Yu-Shiba-Rusinov dimers, Phys. Rev. Lett. **120**, 156803 (2018).

[35] D. Chatzopoulos, D. Cho, K. M. Bastiaans, G. O. Steffensen, D. Bouwmeester, A. Akbari, G. Gu, J. Paaske, B. M. Andersen, and M. P. Allan, Spatially dispersing Yu-Shiba-Rusinov states in the unconventional superconductor FeTe$_{0.55}$Se$_{0.45}$, Nat. Commun. **12**, 298 (2021).

[36] N. Sharma, M. Toole, J. McKenzie, S. Ran, and X. Liu, Atomic-scale frustrated Josephson coupling and multi-condensate visualization in FeSe, Nat. Mater. **24**, 1709 (2025).

[37] C. Castenmiller, R. van Bremen, K. Sotthewes, M. H. Siekman, and H. J. W. Zandvliet, Combined I(V) and dI(V)/dz scanning tunneling spectroscopy, AIP Advances **8**, 075013 (2018).

[38] M. Graham and D. K. Morr, Imaging the spatial form of a superconducting order parameter via Josephson scanning tunneling spectroscopy, Phys. Rev. B **96**, 184501 (2017).

[39] Z. Hou, Z. Shang, W. Duan, W. Xie, H. Yang, H.-H. Wen, Distinct modulation behavior of superconducting coherence peaks associated with sign-reversal gaps in FeTe$_{0.55}$Se$_{0.45}$, arXiv:2512.01703.

[40] X. Dong, F. Zhou, and Z. Zhao, Electronic and superconducting properties of some FeSe-based single crystals and films grown hydrothermally. *Front. Phys.* **8**, 586182 (2020).

[41] X. Liu, and M.C. Hersam, Interface characterization and control of 2D materials and heterostructures. *Adv. Mater.*, **30**, 1801586 (2018).

[42] D.J. Choi, C. Rubio-Verdú, J. De Bruijckere, M.M. Ugeda, N. Lorente, and J.I. Pascual, Mapping the orbital structure of impurity bound states in a superconductor. *Nat. Commun.*, **8**, 15175 (2017).

[43] M. Ruby, B.W. Heinrich, J.I. Pascual, and K.J. Franke, Experimental demonstration of a two-band superconducting state for lead using scanning tunneling spectroscopy. *Phys. Rev. Lett.* **114**, 157001 (2015).

[44] M. Chen, Q. Tang, X. Chen, Q. Gu, H. Yang, Z. Du, X. Zhu, E. Wang, Q.-H. Wang, and H.-H. Wen Direct visualization of sign-reversal s± superconducting gaps in FeTe$_{0.55}$Se$_{0.45}$. Phys. Rev. B, 99, 014507 (2019).

*Corresponding author: xliu33@nd.edu

†These authors contributed equally to this work